\documentstyle[epsfig,aps,pra,twocolumn,floats]{revtex}
\newcommand{\beq}{\begin{equation}}
\newcommand{\enq}{\end{equation}}
\newcommand{\half}{\hbox{$1\over2$}}
\newcommand{\r}{{\bf r}}
\newcommand{\bmu}{{\bf\mu}}
\newcommand{\ie}{{\it i.e.}}

\begin{document}


\draft

\wideabs{

\title{Comment on ``Bose-Einstein condensation with
magnetic dipole-dipole forces''}

\author{J.-P. Martikainen$^1$, Matt Mackie$^1$, and K.-A. 
Suominen$^{1,2}$}

\address{$^1$Helsinki Institute of Physics, PL 9, FIN-00014
        Helsingin yliopisto,Finland\\
        $^2$Department of Applied Physics, University of Turku, FIN-20014,
Turun yliopisto, Finland}

\date{\today}

\maketitle

\begin{abstract}
The ground state solutions of a dilute Bose condensate with contact and
magnetic dipole-dipole interactions are examined. By lowering the value
of the scattering length, Goral {\it et al.} {[}$\,$cond-mat/9907308 and 
Phys. Rev. A {\bf 61}, 051601 (2000)$\,${]} numerically predict 
a region of unstable
solutions,  accompanied by a neighborhood where the ground-state wave
functions have internal structure. On the contrary, we find that the
dipolar condensate has an intuitively-located stability region, and
ground-state solutions near the instability threshold which are absent
any unusual structure.
\end{abstract}
}

\narrowtext

In a recent paper \cite{Goral}, Goral {\it et al.} studied a non-ideal
Bose-Einstein condensate in the presence of magnetic
dipole-dipole forces. For a magnetic moment equal to that for chromium
and a typical value for the scattering length, all solutions were found
stable and differed only in size from condensates lacking long-range
interactions. Upon reducing the scattering length below
a critical value they observed the dipolar condensate to
become unstable, and for scattering length values just above the
instability a ``structured'' condensate was found
(\ie, the density had several peaks
instead of the ordinary single-peaked behavior). The purpose of this
Comment is to report a disagreement in both the location of the
instability threshold, and the existence of the structured
solutions.

The results of Ref.~\cite{Goral} were achieved by numerically solving 
the
Gross-Pitaevskii equation for atoms in a cylindrical harmonic trap:
\begin{eqnarray}
\label{GP}
          i\hbar{\partial\Psi\over\partial t}&=&
          \left\{ -{\hbar^2\nabla^2\over2m}
           +\half m\omega_0^2(x^2+y^2+ \gamma^2z^2)
+{4\pi\hbar^2a\over m}N|\Psi|^2
\right.\nonumber\\
&&\left.
+N\int V(\r-\r')|\Psi(\r')|^2 d^3\r'\right\}\Psi.
\end{eqnarray}
Here $\Psi$ is the condensate wave function, $a$ is the s-wave 
scattering
length, $N$ is the number of atoms, and $m$ is the mass of the atom. 
The
long-range potential is due to the magnetic dipole-dipole interaction 
and
is given by
\beq
\label{dipdippot}
V(\r-\r')=\frac{\mu_0}{4\pi}
        \frac{\bmu_1(\r)\cdot\bmu_2(\r')-3\,
        \bmu_1(\r)\cdot{\bf u}\,\bmu_2(\r')\cdot{\bf u}}{|\r- \r'|^2},
\enq
where ${\bf u}=(\r-\r')/|\r-\r'|$ and $\mu_0$ is the magnetic
permeability of the vacuum.  All the magnetic moments are assumed to
point in the same direction ($z$-direction), \ie,
$\bmu_1=\bmu_2=\mu\hat{z}$. As explained in Ref.~\cite{Goral}, the
Gross-Pitaevskii equation~(\ref{GP}) can be solved
easily enough when one notices that the nonlocal term is a convolution,
and is therefore local in $k$-space. This approach requires a
Fourier transform of the dipole-dipole interaction, given
in the limit of small atomic overlap distance as~\cite{Goral}
\beq
   \label{Ftrans}
         {\cal F}[V(\r)]=-\hbox{$1\over3$}\mu_0\mu^2(1-3\cos^2\alpha),
\enq
where $\alpha$ is the angle between ${\bf k}$ and $\bmu$.

Now, inspecting the formula (\ref{Ftrans}), one thing seems
intuitively clear: the first term will introduce an effective
shift in the scattering length according to
\beq
        U_0=\frac{4\pi\hbar^2aN}{m}-\frac{\mu_0\mu^2N}{3},
\enq
and one would expect instability (roughly) when this becomes negative. 
A
first approximation to the instability threshold is then
\beq
        \bar{a}_c=\frac{m\mu_0\mu^2}{12\pi\hbar^2}.
\enq
Using the chromium parameters $\mu=6\mu_B$, $m=52u$ we get the estimate
($a_{\text{Na}}=2.75\,{\rm nm}$)
\beq
        \bar{a}_c/a_{\text{Na}}\approx 0.291.
\label{ACRIT0}
\enq

This simple estimate is about twice as big as the values given by Goral
{\it et al.}~\cite{Goral} in their Fig.~2,  and it also indicates a
critical scattering length that is independent of particle number.
To a certain extent, the kinetic energy (and possibly the anisotropic 
part
of the dipole-dipole interaction) can stabilize the condensate in a
manner similar to that for condensates with negative scattering 
lengths,
but it seems unlikely that these would cause such a dramatic shift in 
the
critical scattering length.

To determine more accurately when the condensate is expected to
turn unstable, we perform a variational calculation by assuming a trial
wave function for the condensate density~\cite{Yi2000}
\beq
        |\Psi|^2=\frac{1}{\pi^{3/2}\sigma^2\sigma_z}
    \exp\left(-(x^2+y^2)/\sigma^2\right)
    \exp\left(-z^2/\sigma_z^2\right).
\enq
The energy of the condensate is then $E=E_K+E_P+E_{NL}+E_{DD}$,
where
\begin{eqnarray}
E_K&=&\frac{\hbar^2}{2m}\left(\frac{1}{\sigma^2}+\frac{1}{2\sigma_z^2}
        \right),\\
    E_P&=&\frac{m\omega_0^2}{2}\left(\sigma^2
    +\frac{1}{2}\gamma^2\sigma_z^2\right),\\
    E_{NL}&=&\frac{U_0}{(2\pi)^{3/2}\sigma^2\sigma_z}\,,
\end{eqnarray}
and
\begin{eqnarray}
    E_{DD}&=&\frac{\mu_0\mu^2}{\sqrt{2}\pi^{3/2}\sigma^2\sigma_z}
    \nonumber\\
    &&\times\left\{ \frac{1}{2}-\frac{\sigma^2}{3\sigma_z^2}
    \,_2F_1\left(\frac{3}{2},2;
    \frac{5}{2};\left[1-\left(\frac{\sigma}{\sigma_z}
    \right)^2\right]\right)\right\}.
\end{eqnarray}
Here $_2F_1(\alpha,\beta;\zeta;z)$ is a hypergeometric function, and
Eq.~(\ref{Ftrans}) has been used to calculate the anisotropic
contribution ($E_{DD}$) to the energy.  Minimizing this energy 
functional
for a spherically symmetric trap with $\omega_0=(2\pi) 150\,{\rm Hz}$ 
and
$N=300\,000$, we find that the energy becomes unbounded and the system
unstable when $a/a_{\text{Na}}\approx 0.94\,\bar{a}_c$.
Additionally, the critical value is quite insensitive to the number
of particles: for $N=10^7$ it is $a/a_{\text{Na}}\approx
0.95\,\bar{a}_c$. These values are in disagreement with the results of
Ref.~\cite{Goral}.

We also used a trial wave function
\beq
\label{ParaGauss}
        |\Psi|^2=\frac{3}{4\pi z_0\sigma^2}\left[1-\left(\frac{z}{z_0}
\right)^2\right]\exp\left(-(x^2+y^2)/\sigma^2\right).
\enq
At the threshold of instability this trial wave function seems
better justified than the Gaussian trial. Due to the
$\cos^2\alpha$ factor in Eq.~(\ref{Ftrans}), there is
extra repulsion in the $z$-direction, so inverted parabola seems
to be a natural guess. In the $xy$-plane the interaction terms
vanish so a Gaussian profile is expected.
For the function (\ref{ParaGauss}) the kinetic energy diverges so it 
was
left out from the optimization process. Thus this approach
is valid only at the Thomas-Fermi limit, \ie, at large
particle numbers.
We were not able to evaluate all the integrals analytically, so a
numerical  integration was performed. With this wave function
we estimate that for $N=300\,000$ the critical value is
$a/a_{\text{Na}}\approx 0.96\,\bar{a}_c$,
again quite close to the simple estimate given before.

\begin{figure}[htb]
\centerline{\epsfig{file=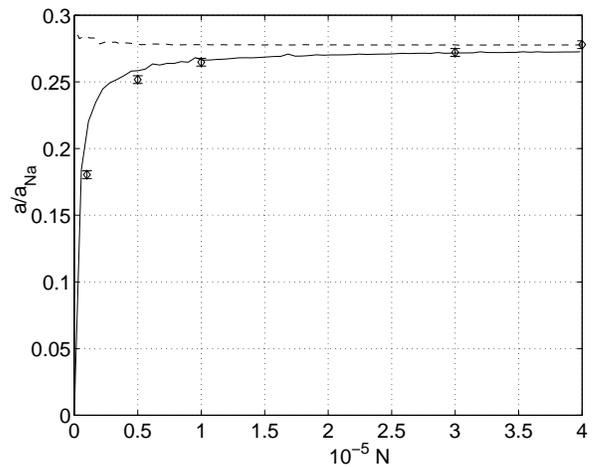,width=7.8cm}}
\vspace*{1cm}
\caption[fig1]{
Instability threshold as a function of the
number of particles for a non-ideal dipolar
condensate of chromium atoms in a spherically symmetric trap with
$\omega_0=(2\pi) 150\,{\rm Hz}$. The solid line is based on a Gaussian trial, and 
the
dashed line is based on a parabolic-Gaussian trial. Individual data
points are numerical solutions of the GP equation with error bars
indicated. (The small amount of noise present in the figure is due to 
the
Monte-Carlo method used for minimization.)
\label{ac_plot}}
\end{figure}

In Fig.~\ref{ac_plot} we show the instability threshold calculated with
three different methods. It can be seen that each method gives
about the same results at large particle numbers. The missing 
kinetic-energy
term for the parabolic-Gaussian trial results in a non-physical 
behavior
at small particle numbers, but the Gaussian trial gives fairly 
satisfactory
results for all particle numbers. Fig.~\ref{ac_plot} should be 
contrasted with
Fig.~2 of Ref.~\cite{Goral}.

Finally, we address the existence of structured dipolar condensates
by solving Eq.~(\ref{GP}) numerically
using the method outlined in Ref.~\cite{Goral}.
The largest grid we used had
a size $128\times 128\times 128$, but the results obtained with
this grid were not very different from those in a grid with a size
$64\times 64\times 64$. For $N=300\,000$ we estimate the critical value
of the scattering length to be
$a/a_{\text{Na}}\approx 0.92\,\bar{a}_c$.
Near the instability threshold, the ground-state
wave functions looked roughly
parabolic along the $z$-axis and Gaussian in the $xy$-plane.
Solutions in the neighborhood of the Ref.~\cite{Goral} instability were
consistently found to be unstable, although {\em transient} structure
was observed en route to collapse when the kinetic energy term was
artificially removed.

While it is true that a variational wave function will not
deliver any more physics than is already present in the trial
wave function, our analytical and numerical results strongly indicate 
that
the numerical instability threshold of  Ref.~\cite{Goral}
is dubious; furthermore, their structured  solutions are not 
numerically
reproducible. In conclusion, we  therefore contend that, upon lowering
the strength of the contact interaction, a non-ideal dipolar BEC 
reveals
an instability threshold which is largely independent of $N$ and given
to a good approximation by the intuitive result~(\ref{ACRIT0}), and
the ground-state wave functions at and near this threshold have
a simple, single-peak profile.

Authors acknowledge the Academy of Finland for financial
support (project 43336).
J.-P. M. acknowledges support from the National Graduate School on
Modern Optics and Photonics.

\end{document}